\documentclass[a4paper,aps,pra,twocolumn,
showpacs,preprintnumbers]{revtex4}
\usepackage{graphics,graphicx,epsfig}
\usepackage[centertags]{amsmath}
\usepackage{amsfonts,times}
\usepackage{amssymb}

\usepackage[centertags]{amsmath}
\usepackage{amsfonts}
\usepackage{amssymb}
\usepackage{amsthm}

\newcommand{\beq}{\begin{equation}}
\newcommand{\eeq}{\end{equation}}
\newcommand{\beqa}{\begin{eqnarray}}
\newcommand{\eeqa}{\end{eqnarray}}

\newcommand{\ket} [1] {\vert #1 \rangle}
\newcommand{\bra} [1] {\langle #1 \vert}

\newcommand{\proj}[1]{\ket{#1}\bra{#1}}
\newcommand{\mean}[1]{\langle #1 \rangle}

\renewcommand{\Re}{\mathrm{Re}}

\begin{document}

\sloppy

\title{Decay and storage of multiparticle entangled states of atoms in  collective thermostat}

\author{A. M.~Basharov \footnote{Permanent address: Laboratory for nonlinear optics, RRC "Kurchatov Institute",
Moscow 123182, Russia. }, V. N.~Gorbachev and A.A.~Rodichkina
\footnote{Electronic address: an@qic.org.ru}}
\affiliation {Laboratory for Quantum Information $\&$ Computation, Aerospace University, St.-Petersburg 190000, Bolshaia Morskaia 67, Russia}

\begin{abstract}

We derive a master equation describing the collective decay of two-level
atoms inside a single mode cavity in the dispersive limit. By considering atomic decay in the collective thermostat, we found a decoherence-free subspace of the multiparticle entangled states of
the $W$-like class. We present a scheme for writing and storing
these states in collective thermostat.

\end{abstract}
\maketitle

\section{Introduction}

When information is encoded in a quantum state of the physical
system, the robustness of the state is an important factor for
successful communication. Due to decoherence, i.e., interaction with the environment, the state of the system can degrade and lose its quantum correlations. One of the possible solution of the decoherence problem is to use decoherence-free subspaces (DFSs); these include wave functions immune to decoherence \cite{Lidar}.

The first DFS has been introduced by Zanardi {\em et al.} \cite{Zanardi} for
two-level atoms interacting with an electromagnetic field playing the role of environment. The wave functions belonging to the DFS are annihilated by the interaction Hamiltonian and, therefore, are left invariant during evolution. Examples of DFSs for various physical systems, in particular, for light, have been
proposed by several authors (see, for example, \cite{1,2,4}).
Weinfurter {\em et al.} have demonstrated experimentally decoherence-free
quantum communication based on the four-photon polarized states
\cite{DFcommun}.

Simple observations show that quantum correlations between particles
can be produced and maintained in collective processes. These are interesting for DFSs and a large number of physical systems with collective interactions can be found. For the Dicke model with
a single resonant mode Bonifatio {\em et al.} have found a master equation
describing a collective atomic decay when illuminated with a resonant
mode  \cite{Bonifatio}. Palma and Knight {\em et al.} \cite{Knight} have shown that
two-atom decay can result in pure entangled states in a collective
squeezed thermostat. However, entanglement of two atoms can be
achieved in collective decay with vacuum thermostat, when atoms are
placed inside a cavity, as it has been shown by Basharov \cite{Askhat}.
There is a simple reason for collective decay in the cavity scheme.
If atoms interact with a single cavity mode, then atomic relaxation
arises because radiation leaves the cavity. This can be modeled as
interaction between the mode and an external broadband field,
which plays role of a thermostat. Therefore, the atoms, being coupled with the
single mode, have a collective decay.

The aim of this paper is to investigate the collective decay of
atoms in the entangled state. To achieve this we consider a simple model of
two-level atoms inside a single mode cavity and a broadband field.
For this model Klimov {\em et al.} \cite{Klimov}
derived master equations for atoms and a collective relaxation
operator in the dispersive limit, assuming a vacuum bath. We
discuss a more general squeezed thermostat, for which a master
equation is derived using a formalism of unitary transformations.
There are some differences between relaxation operators due to
methods of derivation of master equations.
From the physical point of view, in the dispersive limit there is
only an exchange of phase between atoms and cavity mode. This is
described by the effective Hamiltonian we found, which is diagonal over
atomic and field variables. However, the behavior of atoms becomes more
complicated due to an interaction between the cavity mode and broadband
field, which plays the role of a thermostat. Then one finds a coupling
of atoms with the thermostat, providing atomic collective
relaxation. In this aspect our results differ from that of Ref.~\cite{Klimov}.

In contrast to \cite{Klimov} we use the idea developed in Ref. \cite{Accardi}; we first find the effective Hamiltonian of the
total system including the thermostat and then derive the master equation.
Using the master equation we consider the dynamics of a class of
multiparticle entangled states, which is a slightly generalized $W$
class introduced by Cirac {\em et al.} \cite{Cirac}. Some of the optical and atomic
implementations of the presented states have been demonstrated
experimentally by Weinfurter {\em et al.} \cite{Weinf} and Schmidt-Kaler
\cite{Sch-Kalle}. Some properties of these states, different schemes 
to generate them, and several applications have been considered in 
Ref.~\cite{revieWwW}. We also find that in the case when these 
entangled states are reduced to the Dicke states, they belong to DFS
and are immune to collective decay. To explain this feature we
use symmetry arguments. In fact, the total space of the Dicke
states is represented by irreducible subspaces distinguished by their
symmetry type. The collective interaction we consider does not mix the wave
functions from different subspaces due to symmetry conservation. 
Using these properties, we present a model of a quantum memory for 
writing, storing and reading information encoded in these entangled states.

The paper is organized as follows. First we derive the master
equation in the dispersive limit assuming a general model of
thermostat. Then we introduce a set of multiparticle entangled
states which can be reduced to the Dicke family and consider their
decay in squeezed and vacuum thermostats. Finally we present a scheme for
reading and storing entangled states in the collective thermostat.

\section{Initial equations}

By considering the interaction between atoms and a field, one can
obtain a master equation for one of the systems. This equation is also
known as kinetic and often has Lindblad form. It describes
irreversible processes including atomic relaxation, absorption or
amplification of light, and others phenomena which can be reduced to
the Lindblad equations.

\subsection{Hamiltonian}

We consider $n$ two-level atoms inside a high-finesse optical
cavity, a single cavity mode, and a broadband field outside the
cavity. We assume the Hamiltonian of the system has the form
\begin{eqnarray}
H=H_{a}+H_{c}+H_{b}+V_{1}+V_{2},&&
\end{eqnarray}
where the Hamiltonians of free atoms, cavity mode, and broadband field
are, respectively, $H_{a}=\hbar\omega_{0}R_{3}$, $H_{c}=\hbar\omega_{c}
c^{\dagger}c$, $H_{b}=\sum_{\omega}\hbar\omega
b^{\dagger}_{\omega}b_{\omega}$; here
$R_{3}=\sum_{j}(\ket{1}_{j}\bra{1}-\ket{0}_{j}\bra{0})$, $0,1$ 
label the lower and upper levels of atom, and
$c,c^{\dagger},b_{\omega},b^{\dagger}_{\omega}$ are creation and annihilation 
operators for photons of the cavity mode and
broadband field, respectively. The interaction between atoms and cavity mode $V_{1}$
has the form
\begin{eqnarray}
V_{1}= g(c^{\dagger}R_{-}+ cR_{+}),
\end{eqnarray}
where collective atomic operators are given by
$R_{\pm}=\sum_{j}R^{(j)}_{\pm}, R^{(j)}_{+}=[R^{(j)}_{-}]^{\dagger}=
\ket{1}_{j}\bra{0}$. The term $V_{2}$ describes two processes: (1)
an interaction between the broadband field and the cavity mode due
to nonzero transmittance of the output mirror; (2) an interaction
between atoms and the broadband field due to non-ideal sidewalls
of the cavity. It reads
\begin{eqnarray}
\label{V2}
 V_{2}=\sum_{\omega}b_{\omega}[\Gamma_{\omega}c^{\dagger}+
\sum_{j}K_{\omega j}R^{(j)}_{+}]+h.c.&&
\end{eqnarray}
From the physical point of view the broadband field plays the role of a
thermostat and causes the relaxation of atoms and cavity mode.
Relaxation terms can be achieved by switching on this field. It can
be done in different ways using various approximations.

\subsection{Dispersive limit}

We assume that  detuning $\Delta=|\omega_{c}-\omega_{0}|$ is large
and consider the dispersive limit, which can be justified when
\cite{DispLimit}
\begin{eqnarray}
\label{77} |\Delta|\gg ng\sqrt{\mean{c^{\dagger}c}+1}.
\end{eqnarray}
To derive the master equation let us introduce a transformation of
Hamiltonian $H$ given by a time independent unitary operator $S$ 
\begin{eqnarray}
H'=e^{-iS}He^{iS}=-i[S;H]-(1/2)[S; [S; H]]+\dots.
\end{eqnarray}
Using perturbation theory  over interactions $V_{1}$ and $V_{2}$
one finds the operator $S$, from which an effective Hamiltonian
describing the interaction between atoms and cavity mode can be obtained. This
Hamiltonian is diagonal over the field and atomic variables and has
the form
\begin{eqnarray}
  H_{e}=g^{2}\frac{R_{-}R_{+}+cc^{\dagger}2R_{3}}{\hbar
  \Delta}.
\end{eqnarray}
Under this approximation there is another effective Hamiltonian
$H_{g}$ which describes the interaction between atoms and broadband
field
\begin{eqnarray}
  H_{g}=-\frac{g}{\hbar
  \Delta}\sum_{\omega}\Gamma_{\omega}(R_{+}b_{\omega}+R_{-}
  b^{\dagger}_{\omega}).
\end{eqnarray}
In contrast to the second term in $V_{2}$ the obtained Hamiltonian
$H_{g}$ describes the collective interaction of atoms. 
Indeed, in the usual case of the dispersive limit there is no energy exchange between atoms and light, and this is in accordance with the effective Hamiltonian $H_{e}$, which is similar to Ref.~\cite{Klimov}. In the same time the effective Hamiltonian $H_{g}$ shows that atoms and the thermostat field exchange excitations. This is a particular feature of the dispersive limit due to the initial interaction (\ref{V2}). A close analogy is parametric down conversion in transparent nonlinear media, in which the virtual transitions result in an interaction between
photons.

Now we have a problem specified by $H'=H_{a}+H_{b}+H_{c}+H_{e}+H_{g}+V_{2}$, where broadband field can
be considered as a thermostat in a given state.  Assume the thermostat is $\delta$-correlated and its state is squeezed with a center frequency $\Omega$:
\begin{eqnarray}
\label{st}
\nonumber
 \mean{b^{\dagger}_{\omega}b_{\omega'}}=
N(\omega)\delta_{\omega,\omega'},&&\\
\nonumber
 \mean {b_{\omega}b^{\dagger}_{\omega'}}=
(N(\omega)+1)\delta_{\omega,\omega'},&&\\
\nonumber
 \mean{b_{\omega}b_{\omega'}}=
M(\omega)\delta_{2\Omega,\omega+\omega'},&&\\
\mean{b^{\dagger}_{\omega}b^{\dagger}_{\omega'}}=
M^{*}(\omega)\delta_{2\Omega,\omega+\omega'},&&
\end{eqnarray}
where the photon numbers $N(\omega)$ and $M$ are related as
$|M(\omega)|\leq\sqrt{N(\omega)[N(\omega)+1]}$. A physical model of
this thermostat can be represented by the light generated in
parametric down conversion process. Its  simple non-degenerate
version is described by the next Hamiltonian
$H=\sum_{\omega}(k_{\omega}b^{\dagger}_{\Omega
+\omega}b^{\dagger}_{\Omega-\omega} +H.c.)$, where $2\Omega$ is the pump
frequency and $\omega$ belongs to a frequency band $h$
given by phase matching conditions.  The photon numbers $N$ and
$M$ have the form $N(\omega)=\sinh^{2} r_{\omega},
M(\omega)=\exp(i\arg k_{\omega}) \cosh r_{\omega}\sinh r_{\omega}$,
where $r_{\omega}\sim |k_{\omega}|$ is a squeezing parameter. For a
squeezed vacuum $r\ll 1$ and $N\approx 0$, while $M\approx
\exp(i\arg k_{\omega}) r_{\omega}$. The generated light is
broadband if $h$ is much bigger than all representative
frequencies of the problem, like the atomic and cavity mode decay
rates.
More precisely, assume that the width of the squeezed broadband
field given  by  Eqs.~(\ref{st}) is much bigger than the detuning
$\Delta$ as in (\ref{77}). Then following the
standard procedure of replacing a finite bandwidth system with white noise \cite{G} we can make all parameter of squeezed light to be independent from the frequency:  $N(\omega)=N,
M(\omega)=M$. We assume the squeezed thermostat is modeled by a parametric down conversion source.  Then its  bandwidth  is determined by the phase matching conditions,  which can be experimentally varied on a wide range.
The next step is switching on the broadband field. This can be achieved
by several methods based on projection operator techniques, stochastic
differential equations, and others. In any case we need a Markovian approximation to obtain a closed equation. In our case this means that the evolution of the broadband light is given by the free Hamiltonian
$H_{b}$ only and the thermostat parameters $N$ and $M$ are
frequency independent. As a result we find a master
equation for the density matrix $\rho$ of atoms and cavity mode. The
equation includes an effective Hamiltonian $H_{e}$ and relaxation
terms. In the dispersive limit and in the interaction picture the
master equation has the form
\begin{eqnarray}
\label{MEQ}
\dot{\rho}=-(i/\hbar)[H_{e}; \rho]-T\rho,
\end{eqnarray}
where the relaxation operator $T$ includes three terms of the
Lindblad form:
$T=\sum_{j}\mathcal{L}_{j}+\mathcal{L}_{c}+\mathcal{L}_{a}$. The first term describes the independent decay of atoms in the squeezed thermostat. When the atoms have the same coupling constant $K_{\omega
j}=K_{\omega}$ it reads
\begin{eqnarray}
\label{SaR}
 \nonumber
\mathcal{
L}_{j}\rho=(\gamma_{\downarrow}/2)(R^{(j)}_{+}R^{(j)}_{-}\rho-
2R^{(j)}_{-}\rho
R^{(j)}_{+}+\rho R^{(j)}_{+}R^{(j)}_{-})&&\\
\nonumber
+(\gamma_{\uparrow}/2)(R^{(j)}_{-}R^{(j)}_{+}\rho-2R^{(j)}_{+}\rho
R^{(j)}_{-}+\rho R^{(j)}_{-}R^{(j)}_{+})&&\\
\nonumber -2MK^{2}R^{(j)}_{+}\rho R^{(j)}_{+}
-2M^{*}K^{2}R^{(j)}_{-}\rho R^{(j)}_{-},&&
\end{eqnarray}
where the decay rates of atomic levels are denoted by
$\gamma_{\downarrow}=|K|^{2}(N+1)$, $\gamma_{\uparrow}=|K|^{2}N$,
and
$|K|^{2}=\hbar^{-2}\sum_{\omega}|K_{\omega}|^{2}\delta(\omega_{0}-\omega)$.
In free space one finds that $|K|^{2}$ reduces to the well-known
formula for the spontaneous decay rate
$4\omega_{0}^{2}d^{2}/3\hbar c^{3}$. Equation (\ref{SaR}) describes
spontaneous decay of independent atoms in the squeezed thermostat,
for which the transversal decay rate becomes slow because of
squeezing:
$\gamma_{\perp}=(\gamma_{\downarrow}+\gamma_{\uparrow})/2-\Re\{
MK^{2}\}$. The second term of $T$ is the relaxation of the cavity mode
due to photons leaving the cavity, and has the form
\begin{eqnarray}
\nonumber
\mathcal{L}_{c}\rho=|\Gamma|^{2}[(N+1)(c^{\dagger}c\rho-2c\rho
c^{\dagger}+\rho c^{\dagger}c)&&\\
\nonumber
+N(cc^{\dagger}\rho-2c^{\dagger}\rho c+\rho cc^{\dagger})&&\\
\nonumber
+M(cc\rho -2c\rho c +\rho cc)&&\\
M^{*}(c^{\dagger}c^{\dagger}\rho -2c^{\dagger}\rho c^{\dagger} +\rho
c^{\dagger}c^{\dagger})],&&
\end{eqnarray}
where
$|\Gamma|^{2}=\hbar^{-2}\sum_{\omega}|\Gamma_{\omega}|^{2}\delta(\omega_{c}-\omega)$.
If $R$ is a reflectance of the output cavity mirror, then
$|\Gamma|^{2}\to c(1-R)/2L$, where $L$ is length of the cavity. The
collective decay of atoms is represented by the operator
$\mathcal{L}_{a}$:
\begin{eqnarray}
\label{LCR}
 \nonumber
\mathcal{L}_{a}\rho=|\chi|^{2}[(N+1)(R_{+}R_{-}\rho-2R_{-}\rho
R_{+}+\rho R_{+}R_{-} )&&\\
\nonumber
+N(R_{-}R_{+}\rho-2R_{+}\rho R_{-}+\rho R_{-}R_{+})&&\\
\nonumber
 +M(R_{+}R_{+}\rho-2R_{+}\rho R_{+}+\rho R_{+}R_{+})&&\\
+ M^{*}(R_{-}R_{-}\rho-2R_{-}\rho R_{-}+\rho R_{-}R_{-})],&&
\end{eqnarray}
where $|\chi|^{2}=|g\Gamma/\hbar^{2}\Delta|^{2}\tau, \tau=L/c$. As a
result, in the dispersive limit there are three relaxation operators
describing single-particle and collective decay. They have a straightforward  physical meaning and they differ from the relaxation operator in Ref.~\cite{Klimov}, which has cross terms including
products of the collective atomic operators by operators of the
cavity mode.

In order to consider the collective decay of atoms let us introduce the interaction
picture $\rho'=\exp(-i\hbar^{-1}H_{e}t)\rho\,\exp(i\hbar^{-1}H_{e}t)$
and assume the following approximations: Let the first term in $H_{e}$ and
single-particle relaxation be small, $g^{2}R_{-}R_{+}/\hbar\Delta$, 
$\sum_{j}\mathcal{L}_{j}\ll cc^{\dagger}R_{3}/\hbar\Delta$, 
$\mathcal{L}_{a},~\mathcal{L}_{c}$. This is true if
$g^{2}/\hbar|\Delta|,~ \gamma_{\downarrow,\uparrow}\ll
\mean{cc^{\dagger}}g^{2}/|\Delta|,~ |\chi|^{2}n,$ and
$|\Gamma|^{2}\mean{c^{\dagger}c}/n.$ Then we can neglect the difference between $H_{e}$
and $g^{2}2c^{\dagger}cR_{3}/\hbar\Delta$ so that the master equation for
the atomic density matrix $f=\mbox{Tr}_{c}\rho'$  is
\begin{eqnarray}
\label{ME}
 \dot{f}=-\mathcal{L}_{a}f.
\end{eqnarray}

\section{Collective decay and storage of entangled states}

When considering decay of atoms one finds that quantum correlations
between particle can be supported in collective thermostats and the
final or steady state depends from the initial one.

\subsection{Entangled Dicke states}
Let's introduce the multiparticle entangled states, the slight
modification of the $W$ states discovered by Cirac \cite{Cirac}
\begin{eqnarray}
\label{hhh}
  \eta_{n}(1)=q_{1}\ket{10\dots 0}+q_{2}\ket{01\dots 0}+\dots+
  q_{n}\ket{00\dots 1},
\end{eqnarray}
where $\sum_{k}|q_{k}|^{2}=1$. Some of these states are reduced to
the Dicke states $\ket{jma}$ \cite{Dicke}, specified by
three quantum numbers $j,m,a$, where $|m|\leq j=0,\dots, n/2-1,n/2$,
$n$ is a number of particle and parameter $a$ describes the degeneracy
and takes $n_{j}= C_{n}^{n/2+j}-C_{n}^{n/2+j+1}$ values. The numbers
$j$ and $m$ are eigenvalues of two commuting collective operators
$J_{3}$ and $J^{2}=J_{1}^{2}+J_{2}^{2}+J_{3}^{2}$
\begin{eqnarray}
J_{3}\ket{jma}=m\ket{jma},\ J_{2}\ket{jma}=j(j+1)\ket{jma},
\end{eqnarray}
where  $J_{b}$ obeys the commutation relations of the momentum operators 
$[J_{b};J_{c}]=i\varepsilon_{bcd}J_{d}$, $b,c,d=1,2,3$. In
the considered case $J_{1}=(1/2)(R_{-}+R_{+}),
J_{2}=(i/2)(R_{+}-R_{-})$.
 When
\begin{eqnarray}
\label{ZSA}
  \sum_{k}q_{k}=0,
\end{eqnarray}
then there is a set of the zero sum amplitude states
discovered by Pati \cite{Pati}.
However, the wave functions $\eta_{n}(1)$ under condition (\ref{ZSA}) 
belong to the Dicke family  with j=m=n/2-1 \cite{MyJETP}. The states have
the next representation
\begin{eqnarray}
\label{eR}
  \eta_{n}(1)=\sqrt{2}\sum_{k=2}^{n}q_{k}\ket{\Psi^{-}}_{1k}\otimes\ket{0}_{(1k)},
\end{eqnarray}
where $\ket{0}_{(1k)}$ denotes a state of $n-2$ particles (without first and $k$th),
 $\Psi^{-}=(1/\sqrt{2})(\ket{01}-\ket{10})$. Equation
(\ref{eR}) gives the structure of entanglement of $\eta_{n}(1)$; it
tells that one of the particles, say, 1 forms EPR
pairs with all other particles $2,\dots, n$ and this feature is invariant
under permutations of particles. Due to antisymmetric vectors
$\Psi^{-}$ the collective evolution of $n$ particles in the state
(\ref{eR}) involves only $n-2$ particles. This has a simple reason. Two-particle collective operators $R_{\pm}$ and $R_{3}$ annihilate
$\Psi^{-}$, then for any operator of evolution $U$ depending on
$R_{\pm}, R_{3}$ we have
\begin{eqnarray}
\label{UUU}
  U\proj{\eta_{n}}=2\sum_{ks}\ket{\Psi_{1k}^{-}}
  \bra{\Psi_{1s}^{-}}
  U(1k;1s)\ket{0}_{(1k)(1s)}\bra{0},
\end{eqnarray}
where $U(1k;1s)\ket{0}_{(1k)(1s)}\bra{0}$ acts on all 
particles except for $1,k$ and $1,s$. These features allow us to get
simple exact solutions for several problems of collective decay
of $\eta_{n}(1)$.

\subsection{Collective squeezed thermostat}

In squeezed thermostat there is an interesting feature. It can
produce or store quantum correlations between particles for several
initial states.
Considering a two-atom collective decay with initial density matrix 
$f(0)=A\proj{00}+B\proj{\Psi^{+}}+C\ket{11}\bra{00}+C^{*}\ket{00}\bra{11}+D\proj{11}$,
where $\Psi^{+}=(1/\sqrt{2})(\ket{01}+\ket{10})$, $A+B+D=1$, one
finds the next pure steady state \cite{Knight}
\begin{eqnarray}
\label{SST}
 s=(\sqrt{N+1}\ket{00}+\sqrt{N}\ket{11})/(\sqrt{2N+1}).
 \end{eqnarray}
This state is entangled. Note that this solution is correct for the initially
symmetric state $f(0)$.
\\
Consider the collective decay of the two entangled states $\eta_{3}$ and
$\eta_{4}$  described by (\ref{ME}). Under conditions
(\ref{ZSA}) the wave functions read
\begin{eqnarray}
\nonumber
  \eta_{3}=q_{2}\Psi^{-}_{12}\ket{0}_{3}+q_{3}\Psi^{-}_{13}\ket{0}_{2},&&\\
  \eta_{4}=q_{2}\Psi^{-}_{12}\ket{0}_{23}+q_{3}\Psi^{-}_{13}
  \ket{0}_{24}+q_{4}\Psi^{-}_{14}\ket{0}_{23}.&&
\end{eqnarray}
According to Eq.~(\ref{UUU}) the evolution of the density matrix
$\proj{\eta_{3}}$ reduces to the dynamics of the single-particle state
$\proj{0}$ for which there is a simple solution $\proj{0}\to
\lambda\proj{0}+(1-\lambda)\proj{1}$. The $\lambda=N/(2N+1)$ is the
occupation number of the lower atomic level. One finds that
$\eta_{3}$ decays into a mixed state with complex structure. This fact can be
explained using a symmetry argument, which tells that under the
single-particle decay the symmetry of the initial state is not
conserved. In contrast dynamics of  $\eta_{3}$, the dynamics of $\eta_{4}$ has other features.
To obtain the solution we use (\ref{SST}) and find that the final state is
obtained by replacing $\ket{00}\to s$
\begin{eqnarray}
\eta_{4}\to q_{2}\Psi^{-}_{12}\ket{s}_{23}+q_{3}\Psi^{-}_{13}
  \ket{s}_{24}+q_{4}\Psi^{-}_{14}\ket{s}_{23}.
\end{eqnarray}
From this equation it follows that the state is pure, has a more
complicated entanglement structure, but as before one of the atoms forms EPR
pairs with all another atoms.

\subsection{Vacuum thermostat}

Assuming a simple thermostat model for which $M=N=0$, the
master equation (\ref{ME}) reduces to
\begin{eqnarray}
\label{VacT}
  \dot{f}=-\kappa(R_{+}R_{-}f-R_{-}f
R_{+}+fR_{+}R_{-}),
\end{eqnarray}
where $\kappa=|\chi|^{2}$. This equation describes a collective
decay in the vacuum thermostat conserving quantum correlations.
The simplest example is the two-particle antisymmetric function
$\Psi^{-}$ belonging to DFS and immune for decay. The more
interesting examples, introduced by Zanardi \cite{Zanardi}, are DFSs of
multiatom states, products of $\Psi^{-}$.

Suppose the atoms inside the cavity are prepared in the state $\eta_{n}(1)$, then
they evolve according to Eq. (\ref{VacT}) which can be
solved exactly. It is easy to verify that the Lindblad
operator $\mathcal{L}_{0}f=R_{+}R_{-}f-R_{-}FR_{+}+H.c.$ in Eq.~(\ref{VacT}) has the
following properties:
\begin{eqnarray}
\label{LLL}
 \nonumber
\mathcal{L}_{0}\proj{\eta_{n}}&=&Q\ket{1;n}\bra{\eta_{n}}
-|Q|^{2}\proj{0}+Q^{*}\ket{\eta_{n}}\bra{1;n},\\
\nonumber
\mathcal{L}_{0}\ket{\eta_{n}}\bra{1;n}&=&Q\proj{1;n}-2Qn\proj{0}+n
\ket{1;n}\bra{\eta_{n}},\\
\mathcal{L}_{0}\ket{1;n}\bra{\eta_{n}}&=&[
\mathcal{L}_{0}\ket{\eta_{n}}\bra{1;n}]^{\dagger},
\end{eqnarray}
where $Q=\sum_{k}q_{q}$, $\ket{0}=\ket{00\dots 0}$ is a ground state
of atoms and $\ket{1;n}$ is a fully symmetric state,  the
normalized version of which, $W_{n}=(1/\sqrt{n})\ket{1;n}$, is known as $W$ state
\begin{eqnarray}
 W_{n}=(1/\sqrt{n})(\ket{10\dots 0}+\ket{01\dots 0}+\dots
 \ket{00\dots 1}).
\end{eqnarray}
It follows from Eqs. (\ref{LLL}) that the Lindblad operator
$\mathcal{L}_{0}$ maps the set  of states $\{\proj{\eta_{n}}$,
$\proj{0},\proj{1;n}$, $\ket{1;n}\bra{\eta_{n}}$, 
$\ket{\eta_{n}}\bra{1;n}\}$ into itself. This observation allows us to
get an exact solutions for density matrix
\begin{eqnarray}
\nonumber
f(t)=A(t)\ket{1;n}\bra{\eta_{n}}+A^{*}(t)\ket{\eta_{n}}\bra{1;n}
&&\\
 +B(t)\proj{1;n}
 +
 S(t)\proj{0}
 +D\proj{\eta_{n}},
\end{eqnarray}
where  the normalized condition reads
$A(t)Q^{*}+A(t)^{*}Q+B(t)n+S(t)+D(t)=1$ and coefficients obey
equations
\begin{eqnarray*}
\dot{A}=-\kappa(An+DQ),&&\\
\dot{B}=-\kappa(AQ^*+A^*Q)-\gamma nB,&&\\
\dot{S}=-2\kappa(nS+n(1-D)+|Q|^{2}D),&&\\
 \dot{D}=0.&&
\end{eqnarray*}
Similarly to the squeezed thermostat there is a steady state
solution, if $t\to\infty$
\begin{eqnarray}
 \nonumber
f_{ss}=D[-(Q/n)\ket{1;n}+\ket{\eta_{n}}]
[-(Q^{*}/n)\bra{1;n}+\bra{\eta_{n}}]&&\\
+[(1-D)+|Q|^{2}D/n]\proj{0}&&
\end{eqnarray}
which depends on the initial state through parameter $D$. If
$D=0$, then $f_{ss}=\proj{0}$. If $D=1$ one finds evolution of
$\eta_{n}(1)$
\begin{eqnarray}
\label{CDh}
\nonumber
 \proj{\eta_{n}}\to (Q/\sqrt{n})\Big(e^{-n\kappa
t}-1\Big)\ket{W_{n}}\bra{\eta_{n}}+h.c.&&\\
\nonumber
 +(|Q|^{2}/n)\Big(1-e^{-n\kappa t}\Big )^{2}\proj{W_{n}}&&\\
 + (|Q|^{2}/n)\Big(1-e^{-2n\kappa t}\Big)\proj{0}
+\proj{\eta_{n}}.&&
\end{eqnarray}
It follows from (\ref{CDh}) that under condition (\ref{ZSA})
$\eta_{n}(1)$ is Dicke state and has immunity to the collective
decay. However, this result can be obtained without any calculations
because its annihilation by the Lindblad operator
$\mathcal{L}_{0}$. In contrast to $\eta_{n}(1)$, the fully symmetric $W_{n}$ state
degrades: $W_{n}\to \ket{0}$.

The robustness of $\eta_{n}$ entangled states can be clear from the symmetry argument. In the considered collective processes the particle permutation operator is an integral of motion, so that
state symmetry is conserved. Therefore the antisymmetric wave
function $\Psi^{-}$ is robust to decay because the transition
$\Psi^{-}\to \ket{0}$ is forbidden, but the fully symmetric $W$
state can transform into $\ket{0}$. In the case of $\eta_{n}(1)$ the situation is
more complicated, nevertheless symmetry plays a principal role here also.
As it is known the space of Dicke states is represented by irreducible
subspaces distinguished by their symmetry type over particle permutations. Under the condition (\ref{ZSA}) the wave functions
$\eta_{n}(1)$ belong to $[n,n-1]$ irreducible representation of Dicke states in
contrast to $W$ and ground state $\ket{0}$, which belong to  the $[n,0] $
one. Due to symmetry conservation the subspaces of different symmetry
do not mix. This point is in accordance with the fact that in the dynamics
of the wave functions the final states at
$t\to \infty$ are not usual steady states but depend on their subspace and 
initial conditions.
\\
Robustness of $\eta_{n}(1)$ is a natural basis for a quantum memory.
Memory includes writing, storing, and reading of information encoded
by a quantum state. By choosing $\eta_{n}(1)$ states to encode information, a model of quantum memory can be designed. Writing
and reading are achieved by swapping: $a\otimes b \to b\otimes a$. A
particular case of swapping of two mode light in a Fock state into
atomic ensemble has been considered in Ref. \cite{AtEnt}. Here we
introduce a scheme for writing and storing multiparticle
states.
\\
Assume that an interaction between atoms inside cavity and light, presented
by its spacial modes with the wave vectors $j$, has the form
\begin{eqnarray}
  V=i\hbar \sum_{j}f(R_{+}^{j}a_{j}\exp(ijr_{j})-h.c.),
\end{eqnarray}
where $r_{j}$ is a position of $j$th atom. This Hamiltonian describes
an exchange of excitation between a single atom and a single mode.
Assume that initially atoms and field are independent: $\ket{0}_{a}\otimes
\ket{\eta_{n}}_{b}$, where $\ket{0}_{a}$ is the ground state of atoms
and $\ket{\eta_{n}}_{b}$ is the light state given by Eq. (\ref{hhh}),
where $\ket{0},\ket{1}$ are the Fock states with 0 and 1 photons, respectively.
The multimode light can be prepared, i.e., by a set of
beam-splitters, distributing a single photon to different paths.
For simplicity assume $\exp(ijr_{j})\approx 1$, then the evolution
is given by
\begin{eqnarray}
\label{Sw}
 \nonumber
\exp(-i\hbar^{-1}Vt)\ket{0}_{a}\otimes\ket{\eta_{n}}_{b}=
\cos(ft)\ket{0}_{a}\otimes\ket{\eta_{n}}_{b}&&\\
+\sin(ft)\ket{\eta_{n}}_{a}\otimes\ket{0}_{b}.&&
\end{eqnarray}
If $\sin(ft)=1$, the state of light $\ket{\eta_{n}}_{b}$ is swapped
into atoms. Under condition (\ref{ZSA}) it can be stored in the
collective thermostat. Due to the unitarity of transformation (\ref{Sw}), we can achieve a reading of the atomic state.

\section{Conclusions}

Being collective properties of a physical system, quantum correlations
between particles and entanglement can be produced and stored in the
collective processes. These processes can describe
interaction between the physical system and its environment, which
often plays a role of thermostat. In contrast to the usual
thermostat the collective thermostat supports quantum correlations
and it is possible to find a DFS, which is a natural basis for quantum memory.
For the considered example of collective decay atoms inside cavity,
we found that a set of entangled states of the $W$-like class is
decoherence free and therefore is suitable to encode quantum
information for storing it in collective thermostat.

\section{Acknowledgments}

We are grateful to Sergei Kulik for discussion. This work is
partially supported by the Delzell Foundation Inc. and RFBR Grant No.
06-02-16769.

\end{document}